\documentclass[
reprint,
superscriptaddress,
floatfix,
 amsmath,amssymb,
 aps,
 noeprint,nonote,
prb,
]{revtex4-2}

\usepackage{amsmath}
\usepackage{graphicx}
\usepackage{dcolumn}
\usepackage{bm}
\usepackage{hyperref}
\usepackage{braket}
\usepackage{booktabs}
\usepackage[normalem]{ulem}
\usepackage{color}

\newcommand\lyf{LiYF$_{4}$}
\newcommand{\lirefx}[1]{LiY$_{1-x}$#1$_{x}$F$_{4}$}

\newcommand\eg{\textit{e.g.}}
\newcommand\ie{\textit{i.e.}}
\newcommand\cf{\textit{cf.}}
\newcommand{\wn}{$\textrm{cm}^{-1}$}
\newcommand{\chem}[1]{\ensuremath{\mathrm{#1}}}

\newcommand{\ion}[1]{#1$^{3+}$}

\begin{document}

\preprint{APS/123-QED}

\title{Precise determination of low energy electronuclear Hamiltonian for \lirefx{Ho}}

\author{A. Beckert}
\email{adrian.beckert@psi.ch}
\affiliation{Laboratory for Micro and Nanotechnology, Paul Scherrer Institut, CH-5232 Villigen PSI, Switzerland}
\affiliation{Laboratory for Solid State Physics, ETH Zurich, CH-8093 Zurich, Switzerland}
\author{R.~I. Hermans}
\affiliation{London Centre for Nanotechnology, University College London, 17-19 Gordon Street, London WC1H 0AH, United Kingdom}
\author{M. Grimm}
\affiliation{Condensed Matter Theory Group, LSM, NES, Paul Scherrer Institut, CH-5232 Villigen PSI, Switzerland}
\author{J.~R. Freeman}
\author{E.~H. Linfield}
\author{A.~G. Davies}
\affiliation{School of Electronic and Electrical Engineering, University of Leeds, Woodhouse Lane, Leeds LS9 2JT, United Kingdom}
\author{M. M\"uller}
\affiliation{Condensed Matter Theory Group, LSM, NES, Paul Scherrer Institut, CH-5232 Villigen PSI, Switzerland}
\author{H. Sigg}
\author{S. Gerber}
\author{G. Matmon}
\affiliation{Laboratory for Micro and Nanotechnology, Paul Scherrer Institut, CH-5232 Villigen PSI, Switzerland}
\author{G. Aeppli}
\affiliation{Laboratory for Micro and Nanotechnology, Paul Scherrer Institut, CH-5232 Villigen PSI, Switzerland}
\affiliation{Laboratory for Solid State Physics, ETH Zurich, CH-8093 Zurich, Switzerland}
\affiliation{Institute of Physics, EPF Lausanne, CH-1015 Lausanne, Switzerland}

\date{\today}

\begin{abstract}
We use complementary optical spectroscopy methods to directly measure the lowest crystal-field energies of the rare-earth quantum magnet \lirefx{Ho}, including their hyperfine splittings, with more than $10$ times higher resolution than previous work. We are able to observe energy level splittings due to the $^6\mathrm{Li}$ and $^7\mathrm{Li}$ isotopes, as well as non-equidistantly spaced hyperfine transitions originating from dipolar and quadrupolar hyperfine interactions.
We provide refined crystal field parameters and extract the dipolar and quadrupolar hyperfine constants ${A_J=0.02703\pm0.00003}$\,\wn \ and ${B= 0.04 \pm0.01}$\,\wn, respectively. Thereupon we determine all crystal-field energy levels and magnetic moments of the $^5$\textit{I}$_8$ ground state manifold, including the (non-linear) hyperfine corrections. The latter match the measurement-based estimates. The scale of the non-linear hyperfine corrections sets an upper bound for the inhomogeneous line widths that would still allow for unique addressing of a selected hyperfine transition \eg~for quantum information applications. Additionally, we establish the far-infrared, low-temperature refractive index of \lirefx{Ho}.
\end{abstract}

\maketitle

\section{Introduction}
\label{intro}
The quantum magnet \lirefx{Ho} has been shown to exhibit a variety of quantum many-body phenomena, such as quantum annealing \cite{brooke1999}, long-lived coherent oscillations \cite{ghosh2002}, long-range entanglement \cite{ghosh2003}, quantum phase transitions \cite{ronnow2005} and high-$Q$ nonlinear dynamics \cite{silevitch2019}. Importantly, knowledge about the hyperfine (HF) interactions and the lowest-energy crystal-field (CF) states is necessary for the understanding of the material's properties.

In addition, rare-earth doped crystals are promising candidates for quantum information applications \cite{ortu2018,kindem2020,raha2020,grimm2020}. Among the many potential host materials, isotopically-pure \lirefx{Er} exhibits long electronic coherence times~\cite{kukharchyk2018}. To control the electro-nuclear degrees of freedom \eg \ for quantum information processing, precise knowledge of the underlying electronuclear Hamiltonian is required. 

Owing to limitations in resolution and other experimental challenges, comprehensive high-resolution measurements of the transitions within all low CF levels of \lirefx{Ho} have not been available so far~\cite{magarino1980,kjaer1989,babkevich2015,matmon2016}, although we have recently combined optical comb synthesis and a software controlled  modulator to obtain ultra-high resolution data for the HF-split lowest CF excitation near 6.8\,\wn~\cite{hermans2020}. Here we integrate theory together with low-temperature terahertz time-domain spectroscopy (TDS) and synchrotron-based ultra-high resolution Fourier transform infrared (FTIR) spectroscopy (Sec.~\ref{method}) to re-examine the transitions between the three lowest-lying CF levels of the $^5$\textit{I}$_8$ ground-state manifold of \lirefx{Ho}.

Motivated by the precisely and unambiguously resolved HF splitting, we expand previous treatments of the HF interaction in CF states (\eg~\cite{elliot1972,matmon2016}) up to second order in the dipolar HF interaction, and to first order in the quadrupolar coupling in Sec.~\ref{CF transitions and HF}. There, we discuss the measured data, in particular the transition energies including isotopic shifts due to $^6\mathrm{Li}$ and $^7\mathrm{Li}$.
We extract CF parameters by combining our data with CF energy measurements of the $^5$\textit{I}$_8$ manifold from Ref.~\cite{matmon2016}, which enables us to refine the dipolar HF interaction constant $A_J$ in Sec.~\ref{sec: CF parameters}. Based on the resulting CF parameters, we predict all $^5$\textit{I}$_8$ CF energies and their magnetic moments. The high instrumental resolution allows us also to determine the quadrupolar HF constant $B$. Using $B$ and $A_J$, we infer non-equidistant HF corrections of the three CF levels involved in our measurements, including the ground state. We provide an approximation of these HF corrections based on our measurement results, which corroborate our numerical simulation. We conclude in Sec.~\ref{sec: CF parameters} by discussing the implications of non-equidistant HF corrections on unambiguous addressing of specific HF transitions \eg \ for quantum information applications. In the appendix~\ref{Sec: Refractive Index} we provide a refractive index measurement of \lirefx{Ho} from $70$ to $5$\,\wn \ corresponding to $\sim2$~THz to the sub-THz range; these data are useful for planning the design of future optical experiments and devices. A summary and an outlook are found in Sec.~\ref{conclusion}.

\section{Experimental setup}
\label{method}

\subsection{Sample}
For an overview of the physical properties of \lirefx{Ho}, we refer to Refs.~\cite{gingras2011,matmon2016}. We study three commercially-available \lirefx{Ho} single crystals at low doping concentrations of \mbox{$x=1\%$}, $0.1\%$, and $0.01\%$. The crystal dimensions along the light propagation direction are chosen such that transmission is optimized for each $x$. Samples were mounted on the cold finger of a continuous-flow liquid-helium cryostat. The THz light was linearly polarized. The sample was oriented with the crystallographic $c$-axis parallel to the magnetic field component, while the light's propagation direction was perpendicular to $c$. All reported temperatures denote the nominal values on the cryostat cold finger.

\subsection{Experimental methods}
\begin{figure}
\includegraphics[width=\columnwidth]{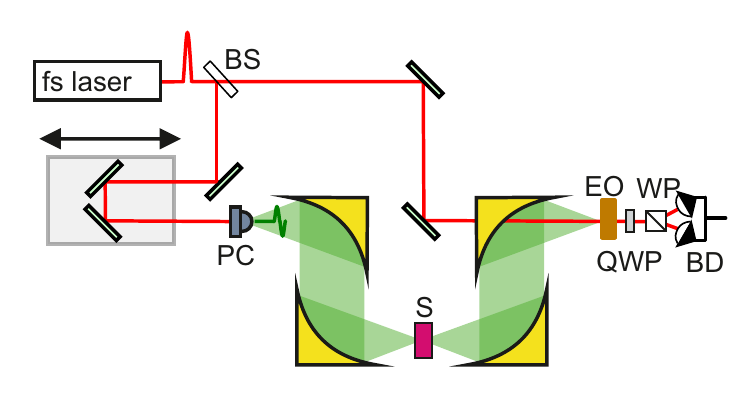}
\caption{Schematic of the time-domain spectroscopy setup showing the femtosecond (fs) laser, the beam-splitter (BS), a photo-conductive (PC) antenna, the sample (S), a ZnTe electro-optic (EO) sampling crystal, a quarter-wave plate (QWP), a Wollaston prism (WP) and balanced photo-diodes (BD) for detection.}
\label{fig:TDSsetup}
\end{figure}

We use two different measurement methods: First, TDS was conducted on \lirefx{Ho} ($x=0.1\%$) for wavenumbers $\tilde{\nu}<10$\,\wn \ (300\,GHz), as well as for refractive-index measurement of the $x=1\%$ crystal for $\tilde{\nu}\le70$\,\wn \  (2.1~THz).
Figure~\ref{fig:TDSsetup} shows a schematic of the custom experimental setup, which is based on an 800\,nm laser, delivering 100\,fs pulses at 80\,MHz repetition rate. The beam is split, directing 250\,mW through a variable delay line. This fraction of the laser is focused onto a low-temperature grown \chem{GaAs} photo-conductive emitter with a 100\,$\mu$m electrode gap (biased at 100\,V, 7.3\,kHz) that generates a linearly polarized single-cycle THz pulse. The THz pulse is then collected from the back of the emitter substrate with a Si hyper-hemispheric lens and focused on the sample with a parabolic mirror. Thereafter, the transmitted beam is refocused onto a 2\,mm thick \chem{ZnTe} crystal for electro-optic sampling. In this detection scheme, the THz branch is overlapped with the 800~nm branch. As a function of delay time, the polarization change of the transmitted 800~nm light is then proportional to the instantaneous THz field in the ZnTe crystal. The signal is measured using balanced photo-diodes and a lock-in amplifier referenced to the emitter bias frequency. Fourier transforms of the delay scans then yield the spectra.

Second, ultra-high resolution FTIR spectroscopy was conducted on \lirefx{Ho} ($x=0.01\%$) for \mbox{$\tilde{\nu}>15$\,\wn}~(450\,GHz) using a custom-built Bruker FTIR spectrometer with 0.00077\,\wn \ (23\,MHz) resolution. A He-flow cryostat for low-temperature measurements was fitted to the spectrometer. The THz source is the high-brilliance, highly collimated far-infrared (FIR) radiation from the Swiss Light Source synchrotron at the Paul Scherrer Institut, Switzerland. Reference~\cite{albert2011b} provides more details about this FTIR setup. The unique combination of a low-temperature, ultra-high resolution spectrometer and FIR synchrotron radiation allowed us to measure the absorbance spectra with a resolution of up to $10^{-3}$\,\wn, which is more than an order of magnitude higher than previously reported ~\cite{karayianis1976,christensen1979,matmon2016}.

The THz response of the holmium ions (\ion{Ho}) in the \lyf \ matrix is characterized by referencing the sample absorption at low temperature to a higher temperature measurement. This ensures that both the background absorption of the crystal host and temperature-independent reflections from the sample and the experimental setup are removed. Therefore, we show absorbance spectra $A(\tilde{\nu})=\mathrm{Log}_{10}[I_0(\tilde{\nu})/I(\tilde{\nu})]$ as a function of wavenumber $\tilde{\nu}$ {[\wn]}, with $I(\tilde{\nu})$ $(I_0(\tilde{\nu}))$ denoting the wavenumber-dependent sample (reference) transmission.

\section{Crystal field transitions with hyperfine interactions}
\label{CF transitions and HF}

Our high-resolution setups enable us to resolve the HF structure of the measured CF states to high precision; analysis methods which take advantage of this structure are described in Ref.~\cite{beckert2020}. We turn now to the theoretical understanding of the HF corrections to the measured CF states. In this paper we denote a transition from an initial CF state $i$ to a final state $f$ by $i\rightarrow f$. Further, we label the $^5$\textit{I}$_8$ ground state manifold states $8.n$ according to their CF energy $E_n$: the ground state ($8.1,E_1=0$) is a doublet (under time-reversal symmetry) and carries $\Gamma_{3,4}$ symmetry, the first excited (8.2) and second excited (8.3) states have $\Gamma_{2}$ symmetry at $E_2=6.8$\,\wn \ and $E_3=23.3$\,\wn, respectively. We denote the CF symmetries (irreducible representations) by $\Gamma_j,~j\in\{1,2,3,4\}$, using standard conventions. 
Individual HF states are labelled as $\ket{8.n^\sigma, m_z} \equiv \ket{8.n^\sigma} \otimes \ket{m_z}$, where $\sigma = -1$ ($\sigma=+1$) denotes the $\Gamma_3$ ($\Gamma_4$) state if the $n$-th level belongs to a doublet. $m_z$ is the nuclear spin projection onto the crystallographic $c$-axis.

\subsection{HF interaction in perturbation theory}
Within the lowest $J$-multiplet, the electrons of each \ion{Ho} ion ($J=8$) couple to their nuclear spin ($I=7/2$) via the dipolar and quadrupolar HF interactions
\begin{equation}
\begin{split}
    H_{\mathrm{HF}} =& H_{\mathrm{HF, dip}} + H_{\mathrm{HF, quad}} \\
    =& A_J \, \vec{J} \cdot \vec{I} + \frac{B}{2I(2I-1)J(2J-1)} \bigg( 3(\vec{J} \cdot \vec{I})^2 \\
    &+ \frac{3}{2} (\vec{J} \cdot \vec{I}) - I(I+1)J(J+1) \bigg),
\end{split}
\label{eq: HF interaction}
\end{equation}
with the dipolar and quadrupolar coupling constants $A_J$ and $B$, respectively. Below we consider the effects of $A_J$ up to second order and $B$ to first order because of the relative size of these terms. We neglect HF corrections due to coupling of the nuclear electric quadrupole moment to the electric field gradient. Using the literature value in Ref.~\cite{popova2000}, this effect is estimated to be an order of magnitude smaller than the terms in Hamiltonian~(\ref{eq: HF interaction}).

Using perturbation theory, the ground-state energy corrections $\delta_{8.1^\sigma, m_z}$ of the states $\ket{8.1^\sigma, m_z}$ are
\begin{equation}
\begin{split}
    &\delta_{8.1^+,+m_z} = \delta_{8.1^-,-m_z}= A \braket{8.1^+|J_z|8.1^+} m_z\\
    &+ \sum_{ j\in \Gamma_{1}}  \frac{A_J^2}{4 \Delta E_{1j}} \bigg[  |\braket{8.j|J_+|8.1^+}|^2 \big( I(I+1)-m_z(m_z+1) \big) \bigg] \\
    &+ \sum_{j \in \Gamma_{2}}  \frac{A_J^2}{4 \Delta E_{1j}} \bigg[  |\braket{8.j|J_-|8.1^+}|^2  \big( I(I+1)-m_z(m_z+1) \big) \bigg]\\
    &+ \sum_{\substack{j \in \Gamma_{3,4}\\ j \neq 1}} \frac{A_J^2}{\Delta E_{1j}} \bigg[ |\braket{8.j^+|J_z|8.1^+}|^2 m_z^2  \bigg]\\
    &+ \frac{B \braket{8.1^+|3J_z^2-J(J+1)|8.1^+}}{4I(2I-1)J(2J-1)}  (3m_z^2-I(I+1)) ,
\end{split}
\label{eq: HF corr doublet}
\end{equation}
and the corrections of the first two excited electronic states ($n=2,3$) are
\begin{equation}
\begin{split}
    &\delta_{8.n, \pm m_z} =  \sum_{\substack{j \in \Gamma_{2}\\ j\neq n}}  \frac{A_J^2}{\Delta E_{nj}}  |\braket{8.j|J_z|8.n}|^2 m_z^2 \\
    &+ \sum_{j \in \Gamma_{3,4}} \frac{A_J^2}{2 \Delta E_{nj}}  \bigg[ |\braket{8.j^+|J_+|8.n}|^2  \big( I(I+1)-m_z(m_z+1) \big) \bigg]\\
    &+ \frac{B \braket{8.n|3J_z^2-J(J+1)|8.n}}{4I(2I-1)J(2J-1)}  (3m_z^2-I(I+1)) .
\end{split}
\label{eq: HF corr singlet}
\end{equation}
Here $\Delta E_{nj}=E_{n}-E_{j}$ is the energy difference between the CF levels $\ket{8.n}$ and $\ket{8.j}$. The sums run over all CF states $\ket{8.j}$ carrying the irreducible representations $\Gamma_i,~i\in\{1,2,3,4\}$. From now on, we use the abbreviation $\lambda_n/2$ for the prefactor of the HF corrections $\propto m_z^2$ of the CF states $8.n$.

These perturbative corrections are sufficient to interpret the HF spectrum of the 8.1, 8.2 and 8.3 states, where $m_z$ denotes the projection of the nuclear spin in the unperturbed electron-nuclear wavefunction. Due to the absence of external magnetic fields, then by Kramers' theorem all HF states are doubly degenerate with their time-reversed state (under time-reversal: $m_z \to -m_z$, $\sigma \to -\sigma$). The electronic doublet 8.1, which is Ising-like with a moment along the crystallographic $c$-axis (due to the $S_4$ site symmetry~\cite{gingras2011}), experiences a dominant first order shift $\propto A_J m_z$ that leads to an equidistant HF splitting into eight HF Kramers doublets. In the lowest ($m_z=-7/2$) and highest ($m_z=+7/2$) of these HF states the electronic and magnetic moments are anti-aligned and aligned, respectively. The singlets do not undergo a first-order HF shift in $A_J$ due to their vanishing moment. Within a single CF state, the equidistance of the HF energies is broken by the second-order terms in $A_J$ and first-order term in $B$, all leading to corrections $\propto m_z^2$. These corrections determine the relative order of the $m_z$ states within a singlet. For the states 8.2 and 8.3, the relative order is reversed. The dominant correction due to the small energy denominator in Eq.~(\ref{eq: HF corr singlet}) comes from the mutual repulsion of the CF states caused by the dipolar HF interaction. An illustration of the HF levels of the 8.2 and 8.3 states is shown in Fig.~\ref{FigTransition82to83}.

\subsection{Experiments}
\begin{figure}[t]
\includegraphics[width=\columnwidth]{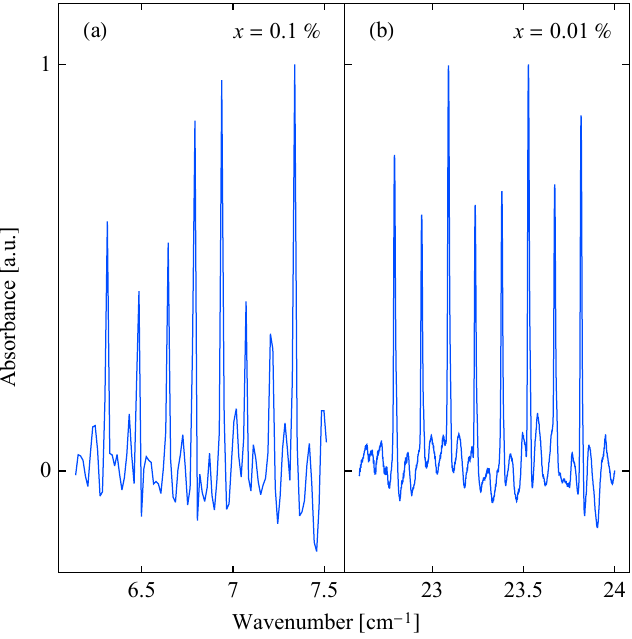}
\caption{Absorbance spectra of \lirefx{Ho} of (a) the $8.1\rightarrow8.2$ (TDS, $x=0.1\%$, $T=3$\,K) and (b) the $8.1\rightarrow8.3$ (FTIR, $x=0.01\%$, $T=3.5$\,K) transitions with conserved $m_z$.}
\label{specplots}
\end{figure}

We measure the transmission of the $8.1\rightarrow8.2$ magnetic dipole transition in \lirefx{Ho} ($x=0.1\%$) at a temperature of $T=2.9$\,K by TDS with an instrument resolution of 0.017\,\wn \ (500\,MHz). The absorbance is shown in Fig.~\ref{specplots}(a), where we directly resolve an eight-fold, approximately equidistant HF splitting of $\sim0.146$\,\wn \, which reflects the dominant linear HF shift of the ground state doublet 8.1. 
The deviation of the individual line intensities from a Boltzmann distribution (\cf \ Refs.~\cite{babkevich2015,matmon2016}) originates from sample- and setup-specific systematic errors such as residual interference of optical components. The extracted Gaussian full-width-at-half-maximum (FWHM) of a single HF line is $0.017\pm0.001$\,\wn \ and thus instrument-resolution limited.

\begin{figure}[ht]
\centering
\includegraphics[width=0.6\columnwidth]{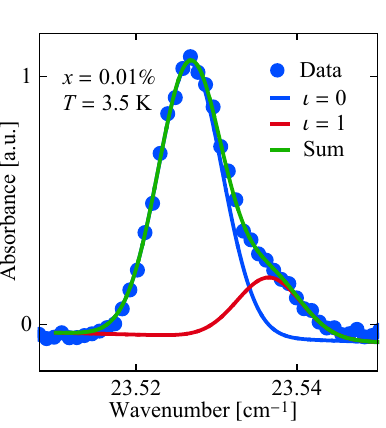}
\caption{Close-up of the asymmetric sixth $8.1\rightarrow8.3$ HF peak of \lirefx{Ho} \mbox{($x=0.01\%$)} at $T=3.5$\,K. The green line is a fit to the main peak (blue), attributed to the majority of Ho ions having $\iota=0$ $^6$Li neighbors, and the smaller second peak (red) that is shifted by isotopic splitting due to $\iota=1$ $^6$Li neighbors. A sinusoidal background owing to interference effects was subtracted w.r.t. the data shown in Fig.~\ref{specplots}.}
\label{detspec}
\end{figure}

The absorbance of the $8.1\rightarrow8.3$ magnetic dipole transition of \lirefx{Ho} \mbox{($x=0.01\%$)} was measured at $T=3.5$\,K with FTIR spectroscopy and 0.001\,\wn (30\,MHz) resolution. The absorbance spectrum is shown in Fig.~\ref{specplots}(b), also revealing the eight-fold CF level splitting. The HF lines are nearly equidistant with a spacing of $\sim0.146$\,\wn. The ultra-high resolution of the FTIR spectrometer allows for a closer inspection of a single HF line. Figure~\ref{detspec} shows the sixth HF peak at $23.527\pm0.001$\,\wn \ in more detail. An asymmetry towards larger wavenumbers is apparent, and is best explained by the isotopic splitting effect due to the natural abundance of $^6$Li ($7.6 \%$) and $^7$Li ($92.4 \%$). A finite number $\iota$ of lighter $^6$Li atoms which substitute the more abundant $^7$Li in the immediate neighborhood of a \ion{Ho} ion was shown to lead to a slight shift in the crystal field parameters of \ion{Ho}, attributed to a difference in the zero point motion of $^6$Li and $^7$Li~\cite{agladze1991,shakurov2005}. This leads to additional peaks in the absorbance spectrum from Ho$^{3+}$ ions with different numbers of less abundant $^6$Li neighbors. Their intensities decrease exponentially with $\iota$~\cite{agladze1991,shakurov2005}, reflecting the Bernoulli distribution of the number of $^6$Li neighbors. We only take the two strongest peaks $\iota=\{0,1\}$ into account, as peaks corresponding to $\iota>1$ were not observed. By fitting two Gaussians, we find an isotopic splitting of $0.0098\pm0.0004$\,\wn \ and a Gaussian FWHM of $0.0090\pm0.0001$\,\wn \ for the individual peaks with the errors extracted from the covariance matrix. These findings are in agreement with the previously reported values of $0.0105\pm0.0015$\,\wn~\cite{shakurov2005}. HF line energies of the $8.1\rightarrow8.3$ transition are always referred to the center of the dominant $\iota=0$ peak.

\begin{figure}[t]
\centering
\includegraphics[width=1\columnwidth]{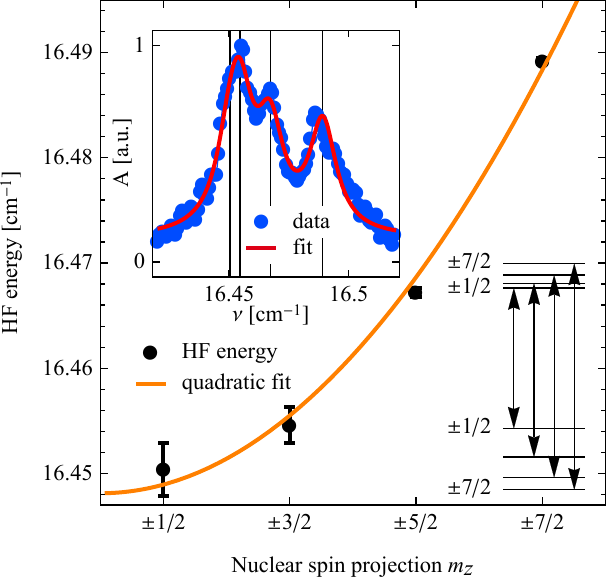}
\caption{\noindent HF shifts of the $8.2\rightarrow8.3$ transition as a function of the nuclear spin  $m_z$. The shifts depend quadratically on $m_z$, as evidenced by the fit (orange). Note the degeneracy of $\pm m_z$ (see Table~\ref{etab}). The top left inset shows the $8.2\rightarrow8.3$ absorbance raw data of \lirefx{Ho} ($x=0.01\%$) at $T=9$\,K (blue) and a fit to Lorentzian profiles (red). The vertical lines denote the peak center positions and correspond to the data in the main figure. Also shown on the bottom right is the energy level diagram of the four observable $8.2\rightarrow8.3$ HF-split CF transitions.}
\label{FigTransition82to83}
\end{figure}

Next we completed FTIR measurements of the $8.2\rightarrow8.3$ magnetic dipole transition of \lirefx{Ho} \mbox{($x=0.01\%$)} with 0.002\,\wn \ resolution. The temperature was set to $T=9$\,K, to thermally populate the 8.2 state. The inset of Fig.~\ref{FigTransition82to83} shows the respective absorbance with a Lorentzian fit. The HF corrections $\propto m_z^2$ in Eq.~(\ref{eq: HF corr singlet}) lead to an observable difference in the transition energies of the individual $m_z$ states. The HF levels are also illustrated (not to scale) in Fig.~\ref{FigTransition82to83}. We fit the absorbance spectrum of the $8.2\rightarrow8.3$ transition with four Lorentzian profiles, taking the degeneracy of $\pm m_z$ into account. We allowed for different intensities and peak frequencies, but imposed an identical linewidth, which we found to be $0.013\pm0.001$\,\wn.
Beyond a 0.008\,\wn \ constant offset, we obtain results that are consistent with the difference measured at the $8.1\rightarrow8.3$ and $8.1\rightarrow8.2$ transitions. We attribute the offset partly to the lower resolution of the TDS setup (0.017\,\wn) and systematic differences between the two experimental setups.

We summarize the HF transition energies in Table~\ref{etab}. Note that the $8.1\rightarrow8.3$ ($x=0.1\%$) transitions, obtained with FTIR, exhibit smaller uncertainties than the $8.1\rightarrow8.2$ ($x=0.01 \%$) transitions, since the latter was measured with lower instrument resolution of the TDS setup. Owing to the significant line overlap of the $8.2\rightarrow8.3$ ($x=0.01 \%$) transition data, the respective uncertainties extracted from the fit covariance matrix amount to $\le 0.003$\,\wn. Reference~\cite{koenz2003} shows that increasing the rare-earth concentrations up to $10\%$ does not noticeably affect the CF energies, which justifies a direct comparison of the $x=0.01\%$ and $0.1\%$ CF energies.

\begin{table}[t]
\begin{tabular}{@{}c|l|ccc@{}}
\toprule
HF Index~~ &~~$m_z$~~ & ~~$8.1^{+}\rightarrow8.2$~~ & ~~$8.1^{+}\rightarrow8.3$~~ & ~~$8.2\rightarrow8.3$ \\ \midrule
1     & $-7/2$ & 7.33 & 23.815 & 16.489 \\
2     & $-5/2$ & 7.21 & 23.671 & 16.467 \\
3     & $-3/2$ & 7.08 & 23.527 & 16.455 \\
4     & $-1/2$ & 6.94 & 23.381 & 16.450 \\
5     & $+1/2$ & 6.80 & 23.235 & 16.450 \\
6     & $+3/2$ & 6.64 & 23.088 & 16.455 \\
7     & $+5/2$ & 6.48 & 22.941 & 16.467 \\
8     & $+7/2$ & 6.31 & 22.794 & 16.489 \\ \bottomrule
\end{tabular}
\caption{HF split transition frequencies for CF level transitions  $8.1\rightarrow8.2$ ($x=0.1\%$), $8.1\rightarrow8.3$ ($x=0.01\%$) and $8.2\rightarrow8.3$ ($x=0.01\%$). All data are given in units of \wn. Uncertainties correspond to $\le\pm0.01$, $\le\pm0.001$ and $\le\pm0.003$\,\wn \ for the $8.1\rightarrow8.2$, $8.1\rightarrow8.3$ and $8.2\rightarrow8.3$ transitions, respectively.}
\label{etab}
\end{table}

\section{Extraction of crystal field parameters and hyperfine interactions}
\label{sec: CF parameters}
The CF parameters of \lirefx{Ho} have been estimated previously based on CF level energies obtained as an average over their HF structure due to the limited resolution~\cite{gifeisman1978,christensen1979,goerllerwalrand1993,shakurov2005,babkevich2015}, or by magnetic susceptibility measurements~\cite{hansen1975,beauvillain1980,ronnow2007}. We improve on those earlier results by including the individually-resolved HF energies of all three CF transitions reported here and supplement these data with results from higher-lying CF states from Ref.~\cite{matmon2016}. We fit the CF parameters and the HF coupling constant $A_J$ simultaneously by numerically calculating the transition energies from the CF Hamiltonian (without HF interaction), as well as the HF splitting to first order in $A_J$. The transition energies are weighted with their measurement errors. This procedure only neglects small corrections to the CF energies due to HF interactions and the linear HF shift in $m_z$ due to second order terms in $A_J$, see Eqs.~(\ref{eq: HF corr doublet},\ref{eq: HF corr singlet}). The refined CF parameters are reported in Table~\ref{tab: cfp}, and we extract the HF coupling constant $A_J=0.02703\pm0.00003$\,\wn \ in agreement with previous estimates in the literature of $A_J=0.0282\pm0.0005$~\cite{magarino1980} and $0.0270\pm0.0003$\,\wn \ \cite{mennega1984}. The error bars of the CF parameters and $A_J$ are computed from the covariance matrix.

\begin{table}
\begin{tabular}{@{}l@{}|r@{}}
\toprule
CF parameter~ & Value [\wn]~~~~               \\ \midrule
~~~~~~$B_2^0$                  & $({-2.66}\pm0.05)\times10^{-1}$ \\
~~~~~~$B_4^0$                  & $({1.68}\pm0.04)\times10^{-3}$  \\
~~~~~~$B_4^{4}$                & $({2.81}\pm0.02)\times10^{-2}$ \\
~~~~~~$B_6^0$                  & $({5.74}\pm0.18)\times10^{-6}$  \\
~~~~~~$B_6^4$                  & $({5.60}\pm0.03)\times10^{-4}$  \\
~~~~~~$B_6^{-4}$               & $({0.00}\pm3.84)\times10^{-3}$  \\ \bottomrule
\end{tabular}
\caption{CF parameters extracted from the transition energy measurements. $B_4^{-4}$ is assumed to be zero~\cite{ronnow2007}.}
\label{tab: cfp}
\end{table}

A comparison of our CF parameter values in Table~\ref{tab: cfp} with previous results shows that we predict smaller values than previously, and we obtain significantly smaller values for $B_2^0$ and $B_4^0$. We attribute these corrections to the inclusion of the HF interaction term (to first order in $A_J$) in the Hamiltonian. In particular, fitting the HF structure allows us to use the magnetic moment of the 8.1 and 8.6 doublets (measured in Ref.~\cite{matmon2016}) as an additional constraint on the CF parameters, which determines the first order HF splitting. With the derived CF parameters we find a considerably ($\sim10 \%$) smaller magnetic moment $\mu/\mu_B= \pm 4.49$ of the 8.6 states than with previous CF parameters. We show the computed CF energies of the $^5$\textit{I}$_8$ manifold and their magnetic moments in Table~\ref{tab: elevels}. Compared to earlier reports, we find $2$-$10\%$ deviations for the predicted energies of the CF levels 8.7 to 8.13.

\begin{table}
\begin{tabular}{@{}l|c l c@{}}
\toprule
CF state~~ & ~Energy~[\wn]~& ~Symmetry  & ~~~$\braket{J_z} = \mu/(g_J \mu_\mathrm{B})$ \\ \midrule
~~~~~8.1  & 0~~~~    &  ~~~~~~$\Gamma_{3,4}$  &  ~\phantom{-}5.40       \\
~~~~~8.2  & 6.84   &  ~~~~~~$\Gamma_{2}$  &        \\
~~~~~8.3  & 23.31   &  ~~~~~~$\Gamma_{2}$  &        \\
~~~~~8.4  & 47.60  &  ~~~~~~$\Gamma_{1}$   &       \\
~~~~~8.5  & 56.92   &  ~~~~~~$\Gamma_{1}$   &       \\
~~~~~8.6  & 72.10 &  ~~~~~~$\Gamma_{3,4}$    &  ~-3.59    \\
~~~~~8.7  & 190.88~ &  ~~~~~~$\Gamma_{1}$   &       \\
~~~~~8.8  & 257.47~ &  ~~~~~~$\Gamma_{3,4}$  &   ~-2.30     \\
~~~~~8.9  & 275.31~ &  ~~~~~~$\Gamma_{2}$ &         \\
~~~~~8.10 & 275.38~ &  ~~~~~~$\Gamma_{1}$  &       \\
~~~~~8.11 & 288.66~ &  ~~~~~~$\Gamma_{1}$   &       \\
~~~~~8.12 & 294.65~  &  ~~~~~~$\Gamma_{3,4}$  &  ~\phantom{-}4.51      \\
~~~~~8.13 & 303.37~ &  ~~~~~~$\Gamma_{2}$   &      \\\bottomrule
\end{tabular}
\caption{Calculated energy levels of \ion{Ho} in \lirefx{Ho} \ based on the CF parameters shown in Table~\ref{tab: cfp}. The last column shows the expectation value $J_z$ of the magnetic $\Gamma_4$ states, which is proportional to their longitudinal magnetic moment $\mu$ and the Land\'e $g$-factor $g_J=5/4$.}
\label{tab: elevels}
\end{table}

\begin{figure}
\centering
\includegraphics[width=1\columnwidth]{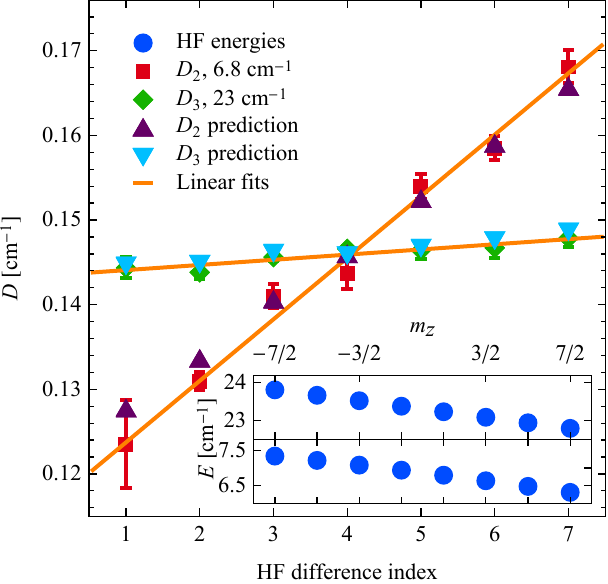}
\caption{\noindent Energy differences $D_{2}$ (red) and $D_3$ (green) between neighboring HF transitions, shown as a function of the HF index. Linear fits to the experimental data are shown in orange. The respective slopes $s_{2,3}$ are a measure of the HF corrections $\propto m_z^2$. The numerical calculations based on our fitted CF parameters are shown in violet and cyan. The HF transition energies are shown in the inset for the $8.1\rightarrow8.2$ and $8.1\rightarrow8.3$ transition in the upper and lower panel, respectively.}
\label{2ndOrderFig}
\end{figure}

In contrast to $A_J$, the determination of the quadrupolar HF interaction constant $B$ requires precise knowledge of the deviations from the linear dipolar HF contributions. We utilize our high-resolution spectra ($8.1\rightarrow8.2$, $8.1\rightarrow8.3$, $8.2\rightarrow8.3$) to fit $B$ separately, by using the determined CF parameters and $A_J$, and numerically calculating the full HF spectrum. We find $B= 0.04 \pm 0.01$\,\wn, which is comparable to the literature value $B= 0.059$\,\wn \ calculated from the free Ho atom~\cite{elliot1972}. 

The parameters $A_J$ and $B$ allow us to numerically compute the HF spectrum. We present a comparison to the experimental data in Fig.~\ref{2ndOrderFig}. To emphasize the HF corrections $\propto m_z^2$, we look at the difference $D_n$ of transition frequencies $8.1\rightarrow8.n$ between neighboring $m_z$ for $n=2,3$. From Eqs.(\ref{eq: HF corr doublet},~\ref{eq: HF corr singlet}), we expect $D_{2,3}$ to be linear in $m_z$, with the slopes being a measure of the HF corrections $\propto m_z^2$. We find the slopes of $D_{2,3}$ (orange lines) to be $s_2=(7.2\pm0.5)\times10^{-3}$\,\wn \ and $s_3=(6\pm1)\times10^{-4}$\,\wn, respectively, based on a linear regression. The numerically calculated values are shown in violet ($D_{2}$) and cyan ($D_{3}$). On average, we find the deviations of the experimental and numerically calculated values to be $16\%$ for the 8.2 and only $1.5\%$ for the 8.3 level, respectively. The error reflects the respective measurement resolutions. 

The experimental values of $D_2$ and $D_3$ allow an order-of-magnitude estimation of the $m_z^2$-correction of the ground state (with the prefactor $\lambda_1/2$). We provide a detailed derivation thereof in the appendix~\ref{app: extraction of HF corr}. Namely, we neglect the quadrupolar interaction $B$ and restrict the sum over the CF states in Eqs.~(\ref{eq: HF corr doublet},~\ref{eq: HF corr singlet}) to the three lowest CF states, which contribute the most to the correction. We then exploit the anti-symmetry of the second-order corrections between the 8.1, 8.2 and 8.3 states to extract $\lambda_1=(s_1+s_2)/4= (2.0\pm0.2)\times10^{-3}$\,\wn \ from our data, \cf \ Eq.~(\ref{eq: HF non-equidistance 1}). This is in agreement with the numerical calculation, yielding $\lambda_1=0.0024$\,\wn. Based on the errors found for the 8.2 and 8.3 energy level predictions, we expect a similar error of $\lesssim16\%$ for $\lambda_{1}$. Akin to $\lambda_1$, we estimate $\lambda_{2}=(-2.5\pm0.1)\times10^{-3}$\,\wn \ and $\lambda_{3}=(1.9\pm0.3)\times10^{-3}$\,\wn \ by also including the $8.2\rightarrow8.3$ transition. Both values are also in agreement with the numerical results $\lambda_{2}=-0.0040$\,\wn \ and $\lambda_{3}=0.0017$\,\wn.

These $m_z^2$-corrections, \ie \ $\lambda_n$, have direct implications on the possibility to unambiguously address HF states, \eg \ in the context of quantum information processing. Specifically, $m_z\to m_z+1$ transitions within an electronic CF state $\ket{8.n}$ can only be driven if the HF line width is smaller than $\lambda_n$, which is the frequency difference of neighboring $m_z\to m_z+1$ transitions. At first sight, the line widths of our spectra seem not to satisfy this criterion. However, a quantitative evaluation of contributions to the line width is necessary to assess whether a regime (temperature, Ho-doping, etc.) for a specific CF state exists, where unambiguous addressing of HF states is possible~\cite{beckert2020c}. This problem can be circumvented by driving protocols involving another excited doublet CF level. Nuclear states can then be manipulated via a first $m_z$-conserving transition to an excited doublet with a subsequent transition to the $m_z$+1 state in the original CF level. For such manipulations involving a nuclear spin flip, the transition energies with different $m_z$ differ already in their first order hyperfine correction and do not rely on the much smaller $m_z^2$-corrections $\lambda$. Our spectra show that this condition is indeed fulfilled.

\section{Conclusions}
\label{conclusion}
We have extended the characterization of the ground CF state manifold of \lirefx{Ho} ($x=1\%$, $0.1\%$, and $0.01\%$) by direct optical measurements of transitions within the lowest three CF states. From the data we calculate the CF parameters, which differ from previous estimates because our refinement also considers the magnetic moments of the CF states as an additional fit constraint via the first order HF shift in $A_J$. In addition, this enables deducing the dipolar HF constant ${A_J=0.02703\pm0.00003}$\,\wn \ purely by optical means. Using the CF parameters we predict the energies for the CF states of the $^5$\textit{I}$_8$ ground state manifold.
Our high measurement resolution allows us to determine the quadrupolar HF constant $B= 0.04\pm0.01$\,\wn and subsequently to calculate the HF corrections of the three lowest CF states. We directly corroborate these calculations via estimations from our data. In addition, we report in appendix~\ref{Sec: Refractive Index} the far-infrared refractive index of \lirefx{Ho}. We conclude that specific addressing of individual HF transitions between doublet states is possible in \lirefx{Ho}, which is important in view of quantum information processing applications~\cite{grimm2020}.

\section{Acknowledgments}
FTIR spectroscopy data was taken at the X01DC beamline of the Swiss Light Source, Paul Scherrer Institut, Villigen, Switzerland. We thank H.~M.~R{\o}nnow, P. Babkevich and J.~Bailey for helpful discussions and experimental support. We thank S.~Stutz for technical support at the X01DC beamline. We acknowledge financial support by the Swiss National Science Foundation, Grant No. 200021\_166271, the European Research Council under the European Union’s Horizon 2020 research and innovation programme HERO (Grant agreement No. 810451), and the Engineering and Physical Sciences Research Council, U.K. (
`HyperTerahertz' EP/P021859/1 and `COTS' EP/J017671/1).

\appendix
\section{Refractive index in the far-infrared}
\label{Sec: Refractive Index}
\begin{figure}[t]
\centering
\includegraphics[width=\columnwidth]{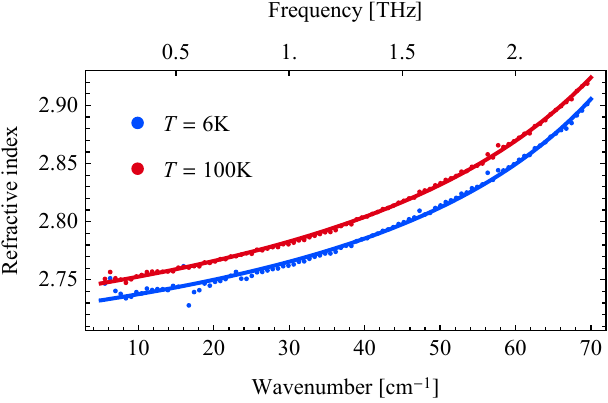}
\caption{\noindent Refractive index of \lirefx{Ho} ($x=1\%$) measured at temperatures of $T=100$ and $6$\,K shown in red and blue, respectively. Solid lines are fits to the data of the phenomenological model described in the main text.}
\label{refindex}
\end{figure}

We report the frequency-dependent refractive index $n$ of \lirefx{Ho} in the FIR regime $10\lesssim\tilde{\nu}\lesssim70$\,\wn. Figure \ref{refindex} shows $n({\nu})$ of a 2.07\,mm thick $x=1\%$ crystal for $T=100$ and 6\,K, as measured with TDS. The results have been obtained from the absorption measurements via the Kramers-Kronig relations, after subtraction of a reference acquired without the crystal in the cryostat. Reflection losses at the sample interfaces have been accounted for.
We fit a phenomenological model $n(\tilde{\nu})=a/(\tilde{\nu}-\tilde{\nu}_0)+c$ to the data, motivated by the divergence of the refractive index near zone-center phonons around $\tilde{\nu}_0=150$\,\wn \ \cite{salauen1997}. From a least squares fit we find $c=2.62\pm0.01$ for both temperatures, $a_\mathrm{6K}=-11.1\pm0.9$\,cm$^{-2}$, $a_\mathrm{100K}=-13.5\pm0.5$\,cm$^{-2}$, $\tilde{\nu}_\mathrm{0,6K}=110\pm2$\,\wn \ and $~\tilde{\nu}_\mathrm{0,100K}=115\pm1$\,\wn.

\section{Hyperfine energies}

\subsection{Perturbation theory}
\label{app: perturbation theory}

The dipolar and quadrupolar HF interaction Hamiltonian is given in Eq.~(\ref{eq: HF interaction}). Rewriting this Hamiltonian in terms of the operators $J_z$, $J_+$, $J_-$ and $I_z$, $I_+$, $I_-$, allows us to derive the perturbative second-order energy corrections in $A_J$ and first-order ones in $B$ as 
\begin{equation}
\begin{split}
    \delta_{8.n^{\sigma}, m_z} &= A_J \braket{8.n^{\sigma}|J_z|8.n^{\sigma}} m_z \\
    &+ \sum_{j\neq i}\sum_{\sigma'=\pm}\frac{A_J^2}{\Delta E_{ij}} \bigg[ |\braket{8.j^{\sigma}|J_z|8.n^\sigma}|^2 m_z^2 \\
    &+ \frac{1}{4} |\braket{8.j^{\sigma'}|J_-|8.n^\sigma}|^2  \big( I(I+1)-m_z(m_z+1) \big) \\
    &+ \frac{1}{4} |\braket{8.j^{\sigma'}|J_+|8.n^\sigma}|^2 \big( I(I+1)-m_z(m_z-1) \big) \bigg] \\
    &+ \frac{B \braket{8.n^\sigma|3J_z^2-J(J+1)|8.n^\sigma}}{4I(2I-1)J(2J-1)}  (3m_z^2-I(I+1)).
\end{split}
\label{eq: HF corr full}
\end{equation}
We have already used here that owing to the $S_4$ crystal symmetry of \lyf, the expectation value of the angular momentum operators with the CF states can only be non-zero for the $J_z$ component, and similarly only the $J_z^2$ component of the quadrupol operators. 

The $S_4$ and time-reversal symmetries simplify the expression~(\ref{eq: HF corr full}) even further, since most of the matrix elements vanish. Due to time-reversal symmetry, the first order correction in $A_J$ is only non-zero for CF doublets, \eg, levels 8.1 and 8.6. Owing to the $S_4$ crystal symmetry (with the symmetry operator being $U= \exp\left(i \frac{\pi}{4} J_z \right)$), the matrix elements $\braket{8.j^{\sigma'}|J_z|8.i^{\sigma}}$ of the second-order corrections in $A_J$ are finite only if the states $\ket{8.i}$ and $\ket{8.j}$ carry the same irreducible representation. Furthermore, $\braket{8.i^{\sigma}|J_+|8.j^{\sigma'}}$ is non-zero only for matrix elements between pairs of states  $\braket{\Gamma_1|J_+|\Gamma_3}$, $\braket{\Gamma_3|J_+|\Gamma_2}$, $\braket{\Gamma_2|J_+|\Gamma_4}$, $\braket{\Gamma_4|J_+|\Gamma_1}$, and---with $i$ and $j$ exchanged---for the Hermitian conjugate matrix elements $\braket{8.i^{\sigma}|J_+|8.j^{\sigma'}}^\dagger=\braket{8.j^{\sigma'}|J_-|8.i^{\sigma}}$ as $J_-=J_+^\dagger$. Here, $\ket{\Gamma_i}$ stands for any CF state that transforms as $\Gamma_i$. Using these symmetry constraints in Eq.~(\ref{eq: HF corr full}), we arrive at Eqs.~(\ref{eq: HF corr doublet},~\ref{eq: HF corr singlet}) in the main text.

\subsection{Extraction of the ground state HF corrections}
\label{app: extraction of HF corr}

In the following we restrict the sum over CF states in Eq.~(\ref{eq: HF corr full}) to the lowest three CF states 8.1, 8.2 and 8.3. This is motivated by the fact that these states give the dominant contributions in the second-order corrections of $A_J$ due to the small energy denominators. Further, we  neglect the quadrupolar coupling $B$, which enables us to estimate the ground state HF energies from our data without prior knowledge of the CF parameters or the constant $A_J$.

Taking into account this reduced Hilbert space of only the three lowest CF states, the energy corrections $\delta_{8.i^\sigma,m_z}$ up to second order in $A_J$ can be written as
\begin{equation}
\begin{split}
    \delta_{8.1^{+},m_z} =& \delta_{8.1^{-},-m_z} \\
    =& K_{1,1}(m_z) + K_{1,2}(m_z) + K_{1,3}(m_z), \\
    \delta_{8.2, \pm m_z} =& K_{2,3}(m_z) + 2 K_{2,1}(m_z),\\
    \delta_{8.3, \pm m_z} =& K_{3,2}(m_z) + 2 K_{3,1}(m_z),\\
\end{split}
\label{eq: HF corr 3states} 
\end{equation}
where $K_{i,j}$ defines the perturbative energy correction of level $i$ due to the level $j$
\begin{equation}
\begin{split}
    K_{1,1}(m_z) =& A_J \braket{8.1^+|J_z|8.1^+} m_z,\\
    K_{1,i=2,3}(m_z) =& \frac{A_J^2}{4}  \frac{|\braket{8.i|J_-|8.1^{+}}|^2}{\Delta E_{1i}}\\
    &\times \big( I(I+1)-m_z(m_z+1) \big),\\
    K_{2,3}(m_z) =& \frac{A_J^2 |\braket{8.3|J_z|8.2}|^2}{\Delta E_{23}} m_z^2,\\
    K_{i,j\neq i}(m_z) =& -K_{j,i}(m_z).
\end{split}
\label{eq: HF corr K}
\end{equation}
Measuring transitions between the 8.1, 8.2 and 8.3 states (with $m_z$ conserved) allows us to extract the second-order ground state HF corrections in $A_J$, \ie \ $K_{1,2}(m_z)+K_{1,3}(m_z)$. We use the anti-symmetry of $K_{i,j\neq i}(m_z)$ in Eq.~(\ref{eq: HF corr K}) to cancel out the contributions $K_{2,3}(m_z)$ in the transition frequencies. We do this by using the differences $D_i(m_z)$ ($i=2,3)$ of transition frequencies $8.1 \to 8.i$ between neighboring $m_z$
\begin{equation}
    \begin{split}
        D_i(m_z)=&(\delta_{8.i,m_z+1}-\delta_{8.1^{+},m_z+1})\\
        &- (\delta_{8.i,m_z} - \delta_{8.1^{+},m_z}).
    \end{split}
\end{equation}
The purely electronic CF transition energies cancel out in $D_i(m_z)$ when we take the difference of two transitions. We add $D_2(m_z)$ and $D_3(m_z)$ to eliminate the contributions $K_{2,3}(m_z)$ and $K_{3,2}(m_z)$ (due to the anti-symmetry of $K$). Taking the difference between neighboring $m_z$, we recover the coefficient of the $\propto m_z^2$ correction in Eqs.~(\ref{eq: HF corr 3states},~\ref{eq: HF corr K}). We introduce $\lambda_1$ which is twice this coefficient: 
\begin{equation}
\begin{split}
    \lambda_1 =& \frac{\mathrm{d} \delta_{8.1^{+},m_z}}{\mathrm{d} m_z}  = \frac{\mathrm{d}}{\mathrm{d} m_z} (K_{1,2}(m_z)+ K_{1,3}(m_z)) \\
    =& -\frac{1}{4} \big[ D_2(m_z+1) + D_3(m_z+1) \\
    &- \left( D_2(m_z) + D_3(m_z) \right) \big].
\end{split}
\label{eq: HF non-equidistance 1}
\end{equation}
The energy difference between neighboring $m_z \to m_z+1$ transitions within the ground state doublet is given by $\lambda_1$. Its value is estimated in the main text by fitting linear functions to $D_i(m_z)$. 

Similarly, we determine the coefficients of the $m_z^2$-HF-correction in the 8.2 and 8.3 states, $\lambda_2$ and $\lambda_3$, respectively, as
\begin{equation}
\begin{split}
    \lambda_2 =& \frac{1}{4} \big[ D_2(m_z+1) + D_3(m_z+1)- 2 D_1(m_z+1) \\
    &- \left( D_2(m_z) + D_3(m_z)- 2 D_2(m_z+1) \right) \big],\\
    \lambda_3 =& \frac{1}{4} \big[ D_2(m_z+1) + D_3(m_z+1)+ 2 D_1(m_z+1) \\
    &- \left( D_2(m_z) + D_3(m_z)+ 2 D_2(m_z+1) \right) \big],
\end{split}
\label{eq: HF non-equidistance 23}
\end{equation}
where we defined the differences $D_1(m_z)$  of transition frequencies $8.2 \to 8.3$ between neighboring $m_z$ as
\begin{equation}
    D_1(m_z)=(\delta_{8.3,m_z+1}-\delta_{8.2,m_z+1})- (\delta_{8.3,m_z} - \delta_{8.2,m_z}).
\end{equation}
\bibliography{AdrianLibrary2020-08-26BibTex}

\end{document}